\documentclass[a4paper,11pt]{article}
\usepackage{mathtext} 
\usepackage[T2A]{fontenc}
\usepackage[utf8]{inputenc}
\usepackage[english]{babel}
\usepackage[left=2cm,right=2cm,top=1.5cm,bottom=1.5cm,bindingoffset=0cm]{geometry} 
\usepackage{amsmath, amssymb} 
\usepackage{mathrsfs} 
\usepackage{graphicx} 
\usepackage{bm} 
\usepackage{dsfont} 
\usepackage{marvosym} 
\usepackage{nicefrac} 

\renewcommand{\phi}{\varphi}

\newcommand{\tr}{\mathrm{tr}}

\newcommand{\pd}{\partial}
\newcommand{\normord}[1]{%
  {:\mathrel{\mspace{2mu}#1\mspace{1mu}}:}%
}

\def\II{\hbox{{1}\kern-.25em\hbox{l}}}

\makeatletter
\renewcommand*\env@matrix[1][\arraystretch]{%
  \edef\arraystretch{#1}%
  \hskip -\arraycolsep
  \let\@ifnextchar\new@ifnextchar
  \array{*\c@MaxMatrixCols c}}
\makeatother

\numberwithin{equation}{section}

\begin{document}



\begin{center}
{\Large \bf{$\mathrm{Q}$-operator for the quantum NLS model}}

\bigskip


{\large \sf N. M. Belousov\,,
S.\,E. Derkachov 
 } 

\vspace{0.5cm}

St. Petersburg Department of Steklov Mathematical Institute of Russian Academy of Sciences, \\
Fontanka 27, 191023 St. Petersburg, Russia

\end{center}


\begin{abstract}
In this paper we show that the operator introduced by A. A. Tsvetkov enjoys all the
needed properties of a $\mathrm{Q}$-operator. It is shown that the $\mathrm{Q}$-operator
of the XXX spin chain with spin $\ell$ turns into Tsvetkov's operator in the
continuum limit as $\ell \rightarrow \infty$.
\end{abstract}

\begin{flushright}
\textbf{Dedicated to M. A. Semenov-Tain-Shansky \\ on the occasion of his 70-th birthday}
\end{flushright}

{\small \tableofcontents}

\section{Introduction}
In the paper~\cite{Z} A. A. Tsvetkov studied an infinite one-dimensional boson system
with pair interaction. The Hamiltonian of the system is given by
\begin{align*}
H = \int \partial_x\bar{\psi}(x)\partial_x\psi(x)\, dx +
\int V(x-y)\bar{\psi}(x)\bar{\psi}(y)\psi(x)\psi(y)\, dx dy,
\end{align*}
and the fields $\bar{\psi}(x)$ and $\psi(x)$ obey the canonical commutation relations
\begin{align*}
\left[\bar{\psi}(x),\bar{\psi}(y)\right] = \left[\psi(x),\psi(y)\right] = 0;
\quad \left[\psi(x),\bar{\psi}(y)\right] = \delta(x-y).
\end{align*}
In the paper~\cite{Z} it was shown that under the certain conditions on the
potential $V(x-y)$ the one-parameter family of operators
\begin{align*}
A(\lambda) = \normord{\exp\left(
\frac{1}{\lambda}\int \bar{\psi}(x)\partial_x\psi(x)\, dx +
\frac{1}{2\lambda^2}
\int V(x-y)\bar{\psi}(x)\bar{\psi}(y)\psi(x)\psi(y)\, dx dy\right)}
\end{align*}
turns out to be commutative $\left[A(\lambda), A(\mu)\right] = 0$. By $\normord{}$
we denote the \textit{normal ordered} form of an operator wherein all creation operators
$\bar{\psi}$ are to the left of all annihilation operators $\psi$. As an example, the potential
$V(x-y) = c\delta(x-y)$ satisfies found conditions. This potential corresponds to
the quantum NLS model.

E. K. Sklyanin proposed that the operator $A(\lambda)$ introduced by
A. A. Tsvetkov is the $\mathrm{Q}$-operator for the
quantum NLS model. The aim of this paper is to prove this conjecture
of E. K. Sklyanin.

The paper is composed of two parts. In the first part we consider
the XXX spin chain of spin $\ell$ with the quantum space carrying
infinite dimensional $s\ell_2$ representations. We discuss in detail
the construction of the local Hamiltonian and the $\mathrm{Q}$-operator
for this model. It is well known~\cite{F1, F3} that the monodromy matrix
and the Hamiltonian of the quantum NLS model can be obtained from
the monodromy matrix and the local Hamiltonian of the XXX spin chain
in the continuum limit as $\ell~\rightarrow~\infty$. Here we investigate the
continuum limit of the $\mathrm{Q}$-operator for the XXX spin chain
constructed in~\cite{D1,D3}. We show that A. A. Tsvetkov's operator emerges
naturally in this limit.

In the second part we derive independently all the needed identities
for the $\mathrm{Q}$-operator of the quantum NLS model.
For the sake of completeness we prove the main statements of the
quantum NLS model with more emphasis on functional methods~\cite{V,B,F} than usual. Also we tried to stress an analogy between the formulas from 
the spin chain and from the NLS model.

\vspace{0.5cm}

\smallskip
{\bf Acknowledgements.}

We are grateful to E. K. Sklyanin for the formulation of the
problem and very useful discussions.
This work is supported by the
Russian Science Foundation (project no.14-11-00598).

\section{XXX spin chain}

\subsection{Monodromy matrix and algebraic Bethe ansatz}

In the XXX spin chain of spin $\ell$ the monodromy matrix is
defined as the product of the $\mathrm{L}$-operators~\cite{STF,S2,F1,F2}
\begin{align}
\label{monodr}
T(u) & =
L_{n}(u) L_{n-1}(u)
\cdots L_{2}(u) L_{1}(u) = \left
(\begin{array}{cc}
A(u) & B(u) \\
C(u) & D(u) \end{array} \right ),
\\
\label{Lax}
L_k(u) & = u +
\vec\sigma\otimes\vec{S}_k = \left (\begin{array}{cc}
u + S_k & S^{-}_k \\
S^{+}_k & u - S_k \end{array} \right ) =
\left (\begin{array}{cc}
u + \ell+a^{\dagger}_k a_k  & -a_k \\
a^{\dagger}_k\left(a^{\dagger}_k a_k +2\ell\right) & u - \ell-a_k^{\dagger} a_k \end{array} \right ).
\end{align}
Each $\mathrm{L}$-operator in equation~\eqref{monodr}
as a matrix acts in auxiliary linear space $\mathbb{C}^2$.
The matrix elements of the operator $L_{k}(u)$
are generators of the Lie algebra $s\ell_2$. They act
in the local quantum space $\mathbb{V}_k$ associated
with the $k$th site. Here we deal with irreducible representation
of the $s\ell_2$ generators parameterized by spin
$\ell$
\begin{equation}\label{Fock}
S_k = a^{\dagger}_k a_k +\ell;
\quad S_k^{-} = -a_k; \quad S_k^{+} =
a^{\dagger}_k\left(a^{\dagger}_k a_k +2\ell\right),
\end{equation}
where $a^{\dagger}_k$, $a_k$ are the creation and annihilation
operators: $a_k\,a^{\dagger}_k - a^{\dagger}_k\,a_k = 1$. Hence
$\mathbb{V}_k$ is a Fock space generated by acting with
the creation operators on a vacuum vector: $a_k|0\rangle = 0$.
In the following we assume that the spin parameter $\ell$ is an
arbitrary complex number. The entries of the monodromy matrix
$T(u)$, the operators $A(u), \ldots,D(u)$, act in the global quantum
space $\mathbb{V} = \mathbb{V}_n\otimes\mathbb{V}_{n-1}\otimes
\cdots\otimes \mathbb{V}_1$.

A convenient and particularly natural choice would be to take
$a^{\dagger}_k = z_k$, $a_k = \partial_{k}$. Therefore $\mathbb{V}_k$
becomes the space of polynomials in one complex variable $\mathbb{C}[z_k]$,
and the vacuum vector is represented by the constant: $|0\rangle =1$.
The global quantum space coincides with the space of polynomials
in $n$ variables $\mathbb{C}[z_1,\ldots,z_n]$, and the entries
of the monodromy matrix are differential operators acting in this
space.

The $\mathrm{L}$-operators satisfy the following local relation
$$
\left(u-v+\mathbb{P}_{12}\right)\left( u +
\vec\sigma_1\otimes\vec{S}_k\right)\left( v +
\vec\sigma_2\otimes\vec{S}_k\right) = \left( v +
\vec\sigma_2\otimes\vec{S}_k\right)\left( u +
\vec\sigma_1\otimes\vec{S}_k\right)\left(u-v+\mathbb{P}_{12}\right).
$$
Here all operators act in a tensor product of three spaces $\mathbb{V}_1\otimes
\mathbb{V}_2\otimes\mathbb{V}_k$, where $\mathbb{V}_1 = \mathbb{C}^2$ and
$\mathbb{V}_2 = \mathbb{C}^2$ are auxiliary spaces, and $\mathbb{V}_k$ is
the local Fock space. By $\mathbb{P}_{12}$ we denote permutation operator:
$\mathbb{P}_{12} \, \vec{x} \otimes \vec{y}~= \vec{y} \otimes \vec{x}$, $\vec{x}
\in \mathbb{V}_1$, $\vec{y}\in \mathbb{V}_2$.

Clearly, from the local relation we can derive the global relation for the monodromy
matrices $T_1(u) = T(u)\otimes\II$ and $T_2(v) = \II\otimes T(v)$
\begin{align}
\label{glob0}
& \left(u-v+\mathbb{P}_{12}\right)\,
T_1(u)\,T_2(v) =
T_2(v)\,T_1(u)\,\left(u-v+\mathbb{P}_{12}\right),
\\
\nonumber
& T_1(u) = \left( u +
\vec\sigma_1\otimes\vec{S}_n\right)\cdots\left( u +
\vec\sigma_1\otimes\vec{S}_1\right); \quad T_2(v) =
\left( v + \vec\sigma_2\otimes\vec{S}_n\right)\cdots\left( v +
\vec\sigma_2\otimes\vec{S}_1\right).
\end{align}
Let us rewrite the relation~\eqref{glob0} in terms of
$4 \times 4$ matrices by using the standard basis
in the tensor product $\mathbb{C}^2 \otimes \mathbb{C}^2$
\begin{align*}
e_1 = |\uparrow\rangle\otimes|\uparrow\rangle, \ \
e_2 = |\uparrow\rangle\otimes|\downarrow\rangle, \ \
e_3 = |\downarrow\rangle\otimes|\uparrow\rangle, \ \
e_4 = |\downarrow\rangle\otimes|\downarrow\rangle; \ \
|\uparrow\rangle = \left(\begin{array}{cc}
1  \\
0 \end{array} \right ), \ \
|\downarrow\rangle = \left(\begin{array}{cc}
0  \\
1 \end{array} \right ).
\end{align*}
With respect to the chosen basis the matrices $T_1(u)$ and
$T_2(v)$ have the form
$$
T_1(u) = \left(\begin{array}{cccc}
A(u) & 0 & B(u) & 0 \\
0 & A(u) & 0 & B(u) \\
C(u) & 0 & D(u) & 0 \\
0 & C(u) & 0 & D(u)
\end{array} \right); \quad
T_2(v) = \left(\begin{array}{cccc}
A(v) & B(v) & 0 & 0 \\
C(v) & D(v) & 0 & 0 \\
0 & 0 & A(v) & B(v) \\
0 & 0 & C(v) & D(v)
\end{array} \right ).
$$
Similarly, the operator $\mathbb{P}_{12}$ is given by
\begin{equation}\label{perm}
\mathbb{P}_{12} = \frac{1}{2}\left(\vec\sigma\otimes \vec\sigma
+\II\right) =
\left(\begin{array}{cccc}
1 & 0 & 0 & 0 \\
0 & 0 & 1 & 0 \\
0 & 1 & 0 & 0 \\
0 & 0 & 0 & 1
\end{array} \right ).
\end{equation}
Substituting these formulas into expression~\eqref{glob0},
we get the system of quadratic relations
for the operators $A(u),\ldots,D(u)$ and $A(v),\ldots,D(v)$.
Note that, as it follows from relation \eqref{glob0}, traces
of the monodromy matrices, transfer matrices $t(u) = A(u)+D(u)$,
commute: $t(u)t(v) = t(v)t(u)$. The eigenvectors of the
transfer matrix $t(u)$ can be found within the framework
of the algebraic Bethe ansatz method~\cite{STF,S2,F1,F2,KBI}.
The following relations are at the heart of this method
\begin{align} \nonumber
C(u) C(v) & = C(v) C(u),\\
\label{AC}
A(u) C(v) & = \frac{u-v+1}{u-v} C(v) A(u)-
\frac{1}{u-v} C(u) A(v),\\
\label{DC}
D(u) C(v) & = \frac{u-v-1}{u-v} C(v) D(u)+
\frac{1}{u-v} C(u) D(v).
\end{align}
The algebraic Bethe ansatz method requires the existence of the
vacuum vector $|0\rangle$ such that
$$
B(u)|0\rangle = 0; \quad  A(u)|0\rangle = \Delta_{+}(u)|0\rangle;
\quad D(u)|0\rangle = \Delta_{-}(u)|0\rangle,
$$
where $\Delta_{\pm}(u)$ are polynomials of degree $n$ in spectral
parameter $u$. In our case, each local quantum space $\mathbb{V}_k$
contains the vacuum vector $|0\rangle_k$. This means that the vector
$|0\rangle_k$ is the eigenvector for the operator $S_k = a^{\dagger}_k a_k+\ell$,
while the operator $a_k$ annihilates it
$$
L_{k}(u)|0\rangle_k = \left (\begin{array}{cc}
u + \ell & 0 \\
\ldots & u - \ell
\end{array} \right )|0\rangle_k.
$$
The global vector $|0\rangle$ is constructed from the local ones:
$|0\rangle = |0\rangle_n\otimes|0\rangle_{n-1}\otimes\cdots \otimes |0\rangle_1$.
Whence we obtain the explicit expressions for the functions $\Delta_{\pm}(u)$:
$\Delta_{\pm}(u) = (u \pm \ell)^n$.

In these terms, the eigenvectors of the transfer matrix have the form
$$
|v_1\ldots v_k\rangle = C(v_1)\cdots C(v_k)|0\rangle; \quad
t(u)|v_1\ldots v_k\rangle = \tau(u)|v_1\ldots v_k\rangle.
$$
All information about parameters $v_i$ is accumulated in
the polynomial
$$
q(u)=(u-v_1)\cdots(u-v_k).
$$
Using~\eqref{AC} and~\eqref{DC} it can be shown~\cite{STF,S2,F1,F2}
that the vector $|v_1\ldots v_k\rangle$ is the eigenvector
of the transfer matrix $t(u)$ with the eigenvalue
\begin{align}\label{tau}
\tau(u)= \Delta_{+}(u) \frac{q(u+1)}{q(u)} + \Delta_{-}(u)
\frac{q(u-1)}{q(u)}
\end{align}
if the parameters $v_i$ obey the Bethe equations:
$$
\Delta_{+}(v_i) q(v_i+1) + \Delta_{-}(v_i)q(v_i-1) = 0;
\quad i = 1,2, \ldots, k.
$$
The Bethe equations have a simple interpretation: $\tau(u)$ is a polynomial,
but each term on the right-hand side of the formula~\eqref{tau} has a pole
at the point $u=v_i$. Thus the residues of these terms should cancel each other.

\subsection{Local Hamiltonian}
To construct the local Hamiltonian, we need the $\mathrm{R}$-operator
that interchanges spectral parameters in the product of
$\mathrm{L}$-operators~\cite{S2}
\begin{align}\label{uv}
R_{21}(u-v) L_2(u) L_1(v) =
L_2(v) L_1(u) R_{21}(u-v).
\end{align}
The $\mathrm{R}$-operator can be represented in the
following form~\cite{D2} ($z_{ik}\equiv z_i-z_k$)
\begin{equation*}
R(u) = \frac{\Gamma(z_{12}\partial_1+2\ell)}{\Gamma(z_{12}\partial_1-u+2\ell)}
\ \ \frac{\Gamma(z_{21}\partial_2+u+2\ell)}{\Gamma(z_{21}\partial_2+2\ell)}.
\end{equation*}
Calculating the derivatives over $u$ of the both sides
of expression~\eqref{uv} and putting $v=u$, we get
$$
R^{\prime}_{21}(0) L_2(u) L_{1}(u) +
R_{21}(0) L_{1}(u) = L_{2}(u) R_{21}(0)+
L_{2}(u) L_{1}(u) R^{\prime}_{21}(0).
$$
At the same time from the explicit formula for the
$\mathrm{R}$-operator we obtain
$$
R_{21}(0) = \II; \quad
R^{\prime}_{21}(0) = \psi(z_{12}\partial_1+2\ell)+
\psi(z_{21}\partial_2+2\ell),
$$
where $\psi(x)$ is the logarithmic derivative
of the Euler gamma function. Finally, we obtain the commutation
relation~\cite{S2}
\begin{equation*}
\left[R^{\prime}_{21}(0),
L_2(u) L_{1}(u)\right] = L_{2}(u) - L_{1}(u).
\end{equation*}
Let us define the operator $H = R^{\prime}_{1 n}(0)+R^{\prime}_{n n-1}(0)+\ldots+
R^{\prime}_{21}(0)$ acting in the global quantum space $\mathbb{V}$.
Therefore from the last expression it follows that this
operator commutes with the transfer matrix $t(u)$.
Here we assume that periodic boundary condition holds
$k+n \equiv k$. Nevertheless, it is particularly convenient
to work with slightly changed operator
\begin{equation*}
H = \sum_{k=1}^{n} H_{k+1k}; \quad
H_{k+1k} = \psi(z_{k k+1}\partial_k+2\ell)+
\psi(z_{k+1 k}\partial_{k+1}+2\ell) - 2\psi(2\ell).
\end{equation*}
In this notation, the vacuum state $|0\rangle$ corresponds
to the zero eigenvalue of the pair Hamiltonian $H_{k+1 k}|0\rangle = 0$.
Let us remark that this Hamiltonian is local. In other words, each term
in the sum describes the interaction between the nearest neighbors only.

\subsection{$\mathrm{Q}$-operator}

Recall that all information about the eigenvector $|v_1\ldots v_k\rangle$
is accumulated in the polynomial $q(u)=(u-v_1)\cdots(u-v_k)$.
Also, the eigenvalues of the transfer matrix are expressed in terms of
the function $q(u)$. Baxter~\cite{B} suggested to interpret the
polynomial $q(u)$ as an eigenvalue of some operator, what we
now call the $\mathrm{Q}$-operator. In the general case,
\begin{align*}
Q(u) |v_1\ldots v_k\rangle = q(u) c(v) |v_1\ldots v_k\rangle,
\end{align*}
where the constant $c(v)$ depends on parameters $v_1, \ldots, v_k$
and doesn't depend on spectral parameter $u$. Usually there exists
the special point $u_0$ such that the $\mathrm{Q}$-operator turns into
the identity operator at this point: $Q(u_0) = \II$. Therefore $c(v) = q^{-1}(u_0)$
\begin{align*}
Q(u) |v_1\ldots v_k\rangle = \frac{(u-v_1)\cdots(u-v_k)}{(u_0-v_1)\cdots(u_0-v_k)}
|v_1\ldots v_k\rangle.
\end{align*}
By definition the $\mathrm{Q}$-operator satisfies the following properties
\begin{itemize}
\item commutativity
$$ [ Q(u) , Q(v)] = 0, \quad
[Q(u), t(v)] = 0;
$$
\item finite-difference Baxter equation
\begin{equation}\label{Baxter equation}
t(u) Q(u) = \Delta_{+}(u) Q(u+1) +
\Delta_{-}(u) Q(u-1).
\end{equation}
\end{itemize}
These properties are sufficient to reproduce all formulas
with the polynomial $q(u)$ obtained in the framework of the
algebraic Bethe ansatz. So, the $\mathrm{Q}$-operator provides
another approach to solve the model. Below we construct the
$\mathrm{Q}$-operator for the considered case~\cite{D1,D3} using 
the most straightforward approach.

\subsubsection{$\mathrm{R}$-operator}

In the representation~\eqref{Fock} the $\mathrm{L}$-operator
has a convenient factorized form
\begin{equation}
L(u_1,u_2) = \left(%
\begin{array}{cc}
  1 & 0 \\
  z & 1 \\
\end{array}%
\right)\left(%
\begin{array}{cc}
  u_1  & -\partial \\
  0 & u_2 \\
\end{array}%
\right)\left(%
\begin{array}{cc}
  1 & 0 \\
  -z & 1 \\
\end{array}%
\right); \quad u_1 = u+\ell-1, \quad u_2 = u-\ell.
\label{LaxFact}
\end{equation}
The main building block in the construction of the
$\mathrm{Q}$-operator is an $\mathrm{R}$-operator
defined by the relation
\begin{equation}
R_{21}(u_1,u_2|v_2)
L_{2}(u_1,u_2)L_{1}(v_1,v_2)=
L_{2}(u_1,v_2)L_{1}(v_1,u_2)
R_{21}(u_1,u_2|v_2), \label{R2}
\end{equation}
where
$$
u_1 = u+\ell-1, \quad u_2 = u-\ell;
\quad v_1 = v+\ell-1, \quad v_2 = v-\ell.
$$
Notice that this $\mathrm{R}$-operator differs from the
operator defined by~\eqref{uv} (the old one interchanges
parameters $u$ and $v$).

The $\mathrm{R}$-operator acts in the space of polynomials
in variables $z_1$ and $z_2$ and could be represented as a
function of operator argument~\cite{D2}
$$
R_{21}(u_1,u_2|v_2) =
\frac{\Gamma(z_{21}\partial_2+u_1-v_2+1)}{\Gamma(z_{21}\partial_2+u_1-u_2+1)}
$$
or as an integral operator
\begin{equation*}
R_{21}(u_1,u_2|v_2) \Phi(z_2,z_1) =\frac{1}{\Gamma(v_2-u_2)}
\int_{0}^{1} d\alpha \, \alpha^{u_1-v_2}
(1-\alpha)^{v_2-u_2-1}
\Phi\left(\alpha z_{21}+z_1, z_1\right).
\end{equation*}
Two forms of the $\mathrm{R}$-operator are connected through
the integral representation of the Euler beta function
$$
B(z_{21}\partial_{2}+a,b-a) =
\int_{0}^{1} d\alpha \, \alpha^{a-1+z_{21}\partial_{2}}
(1-\alpha)^{b-a-1}
$$
and the explicit expression for the acting of the operator
$\alpha^{z_{21}\partial_{2}} = \mathrm{e}^{-z_1\partial_{2}}\alpha^{z_2\partial_{2}}
\mathrm{e}^{z_1\partial_{2}}$:
$$
\mathrm{e}^{-z_1\partial_{2}}\alpha^{z_2\partial_{2}}
\mathrm{e}^{z_1\partial_{2}}\Phi(z_2, z_1) =
\mathrm{e}^{-z_1\partial_{2}}\alpha^{z_2\partial_{2}}
\Phi(z_2+z_1, z_1) = \mathrm{e}^{-z_1\partial_{2}}
\Phi(\alpha z_2+z_1, z_1) = \Phi\left(\alpha z_{21}+z_1, z_1\right).
$$
In our case, $a = u_1-v_2+1$, $b = u_1-u_2+1$.

\subsubsection{Baxter equation}

On the left-hand side of the Baxter equation~\eqref{Baxter equation}
there is a product of the transfer matrix $t(u)$ and the
$\mathrm{Q}$-operator. The building blocks of the transfer
matrix $t(u)$ are the $\mathrm{L}$-operators. Similarly,
we will construct the $\mathrm{Q}$-operator using
the $\mathrm{R}$-operators as its building blocks.
The global relation involving the $\mathrm{Q}$-operator
and the transfer matrix will be derived from the corresponding
local relation for their building blocks. The needed local relation
is the defining equation for the $\mathrm{R}$-operator~\eqref{R2}.
Using the factorized form~\eqref{LaxFact}
of $L_2(u_1,u_2)$ and $L_1(v_1,v_2)$
and the commutativity $\left[R_{21}, z_1\right] = 0$, let us rewrite~\eqref{R2}
$$
Z_2^{-1}
R_{21}(u-v_2)L_2(u_1,u_2)Z_1
= \left(%
\begin{array}{cc}
  u_1 & -\partial_{2} \\
  0 & v_2 \\
\end{array}%
\right)Z_2^{-1}Z_1
\left(%
\begin{array}{cc}
  v_1 & -\partial_{1} \\
  0 & u_2 \\
\end{array}%
\right)R_{21}(u-v_2)\left(%
\begin{array}{cc}
  v_1 & -\partial_{1} \\
  0 & v_2 \\
\end{array}%
\right)^{-1}.
$$
Note that the dependence of the $\mathrm{R}$-operator on the parameter
$v_2$ enters by a shift of the spectral parameter $u$, and
$Z_k \equiv \left(%
\begin{array}{cc}
  1 & 0 \\
  z_k & 1 \\
\end{array}%
\right)$. Computing the product of the matrices on the right-hand side,
we get
$$
Z_2^{-1}R_{21}(u-v_2)L_2(u_1,u_2)Z_1
$$
$$
= \left(%
\begin{array}{cc}
R_{21}(u+1-v_2) +
v_2 R_{21}(u-v_2) & -R_{21}(u-v_2)\partial_{2} \\
-v_2 z_{21} R_{21}(u-v_2) & (u_1-v_2)(u_2-v_2)R_{21}(u-1-v_2) +
v_2 R_{21}(u-v_2) \\
\end{array}%
\right).
$$
Clearly, at the point $v_2 = 0$ the matrix on the right-hand side
becomes upper triangular. So, we put $v_2 = 0$ in the derived matrix relation
and slightly change it by choosing the second space to be the local quantum space 
at site $k$ and the first space to be the local quantum space at site $(k - 1)$
$$
Z^{-1}_{k}R_{k k-1}(u)
L_k(u_1,u_2) Z_{k-1} =
\left(%
\begin{array}{cc}
R_{k k-1}(u+1) & -R_{k k-1}(u)\partial_{k-1} \\
 0 & u_1 u_2 R_{k k-1}(u-1) \\
\end{array}%
\right).
$$
Actually this particular local relation leads to the Baxter equation. Let's turn
to the global objects: appending one additional site $z_0$, we take the product
over all sites
$$
Z_n^{-1} R_{n n-1}(u) \cdots
R_{21}(u) R_{10}(u)
L_n(u)\cdots L_1(u) Z_0
$$
$$
=
\left(%
\begin{array}{cc}
R_{n n-1}(u+1) & -R_{n n-1}(u)\partial_{n} \\
 0 & u_1 u_2 R_{n n-1}(u-1) \\
\end{array}%
\right)\cdots\left(%
\begin{array}{cc}
R_{10}(u+1) & -R_{10}(u)\partial_{1} \\
 0 & u_1 u_2 R_{10}(u-1) \\
\end{array}%
\right).
$$
In this product matrices $Z_k$ and $Z^{-1}_k$ ($k=1,2,\ldots,n-1$)
become neighbors and therefore cancel each other.
Then we calculate the trace over the two-dimensional space
$\mathbb{C}^2$, use commutativity of all $\mathrm{R}$-operators
and $L_k$ with $z_0$ in order to move $Z_0$ to the left and finally
identify sites 0 and $n$. Thus we obtain the Baxter equation
\begin{equation*}
t(u)Q(u) = Q(u+1)+
(u_1 u_2)^n Q(u-1)
\end{equation*}
for operator
\begin{equation*}
Q(u) = \left.R_{n n-1}(u)
\cdots R_{21}(u)R_{10}(u)\right|_{z_0 \to z_n}.
\end{equation*}
It is also possible to check the commutativity properties of
$Q(u)$~\cite{D1,D3}. Now, in order to visualize the
constructed operator we will derive the explicit
formula for the action of $Q(u)$ on polynomials~\cite{D3}.

\subsubsection{Explicit formula for the action on polynomials}

Let us derive a very simple formula for the action
of $Q(u)$ on the global generating function
$(1-x_{n} z_{n})^{-2\ell}\cdots(1-x_{1} z_{1})^{-2\ell}$.
This formula contains in transparent form all information
about the action of the operator $Q(u)$ on polynomials.
The global problem reduces to the local one
\begin{equation*}
R_{n n-1} (u) \cdots
R_{10} (u) (1-x_{n}z_{n})^{-2\ell}\cdots(1-x_{1} z_{1})^{-2\ell}
\end{equation*}
$$
= R_{n n-1}(u)(1-x_{n}z_{n})^{-2\ell} \cdots
R_{21}(u)(1-x_{2} z_{2})^{-2\ell}
R_{10}(u)(1-x_{1} z_{1})^{-2\ell}.
$$
The explicit formula for the action of the $\mathrm{R}$-operator on
a local generating function
\begin{equation*}
R_{k k-1}(u)(1-x_k z_k)^{-2\ell} =
\frac{\Gamma(\ell+u)}{\Gamma(2\ell)} (1-x_k z_k)^{-\ell-u}
(1- x_k z_{k-1})^{-\ell+u}
\end{equation*}
could be obtained using Feynman's formula
$$
\int_{0}^{1}
d\alpha \, \alpha^{a-1} (1-\alpha)^{b-1}
\frac{1}
{\bigl[\alpha A+(1-\alpha)B\bigr]^{a+b}} =
\frac{\Gamma(a)\Gamma(b)}{\Gamma(a+b)} \frac{1}{A^a
B^b},
$$
so that the action on the global generating function is
$$
R_{n n-1} (u)\cdots
R_{1 0} (u) (1-x_{n}z_{n})^{-2\ell}
\cdots(1-x_{1} z_{1})^{-2\ell} =
$$
$$
= \frac{\Gamma^n(\ell+u)}{\Gamma^n(2\ell)}
(1-x_{n}z_{n})^{-\ell-u}(1-x_{n} z_{n-1})^{-\ell+u}\cdots
(1-x_{1}z_{1})^{-\ell-u}(1-x_{1} z_{0})^{-\ell+u}.
$$
Finally, we put $z_0 \to z_n$ and change the norm of the
$\mathrm{Q}$-operator
\begin{align*}
Q(u) \to \frac{\Gamma^n(2\ell)}{\Gamma^n(\ell+u)} Q(u).
\end{align*}
The action of the renormalized operator looks very simple
\begin{equation}\label{Q}
Q(u) \colon (1-x_{n} z_{n})^{-2\ell}\cdots(1-x_{1}z_{1})^{-2\ell} \mapsto
\end{equation}
$$
\mapsto (1-x_{n} z_{n})^{-\ell-u}(1-x_{n} z_{n-1})^{-\ell+u}
\cdots(1-x_{1} z_{1})^{-\ell-u}(1-x_{1} z_{n})^{-\ell+u} .
$$
The operator $Q(u)$ maps any monomial $z_1^{k_1}\cdots z_n^{k_n}$
to polynomial with respect to variables $z_1,\ldots,z_n$ and the
spectral parameter $u$
$$
Q(u) \colon \mathbb{C}[z_1\ldots z_n] \mapsto
\mathbb{C}[u, z_1 \ldots z_n] .
$$
This property guarantees that eigenvalues of $Q(u)$ are
polynomials in $u$. Using formula~\eqref{Q}, we get
$Q(\ell)=\II$, so that we can interpret the polynomial
$\frac{q(u)}{q(\ell)}=\frac{(u-v_1)\cdots(u-v_k)}{(\ell-v_1)\cdots(\ell-v_k)}$
as the eigenvalue of the operator $Q(u)$. Finally, we
obtain the canonical form of the Baxter equation for
the renormalized operator
$$
t(u)Q(u) = (u+\ell)^n Q(u+1)+
(u-\ell)^n Q(u-1).
$$
So, the whole construction of the $\mathrm{Q}$-operator
is based on the local relations.

\section{Continuum limit}

\subsection{Monodromy matrix in the continuum limit}

The monodromy matrix of the continuous model
satisfies the differential equation~\eqref{T0}.
In order to formulate a natural recipe for taking
the $\ell \to \infty$ limit~\cite{F1,KBI,N}, let us
derive a finite-difference version of this differential
equation. First we extract the matrix $\ell\sigma_3$
from the $\mathrm{L}$-operator~\eqref{Lax}
\begin{align*}
L_k(u) = \left(\II+\ell_k(u)\right) \ell\sigma_3;
\quad \ell_k(u) =
\left (\begin{array}{cc}
\frac{u}{\ell}+\frac{1}{\ell} a^{\dagger}_k a_k  & \frac{1}{\ell} a_k \\
2a^{\dagger}_k+ \frac{1}{\ell} a^{\dagger 2}_k a_k &
- \frac{u}{\ell}+\frac{1}{\ell} a^{\dagger}_k a_k  \end{array} \right )
\end{align*}
and consider the difference of the operators
\begin{align*}
T_n(u)- \ell\sigma_3 T_{n-1}(u) = \left(L_n(u)- \ell\sigma_3\right) T_{n-1}(u) = \ell_n(u) \ell\sigma_3 T_{n-1}(u).
\end{align*}
Hence for the following operator
\begin{align*}
\mathbb{T}_n(u) = \left(\ell\sigma_3\right)^{-n} T_{n}(u)
\end{align*}
we obtain a finite-difference equation of the form
\begin{align}\label{diskr}
\mathbb{T}_n(u)-\mathbb{T}_{n-1}(u) &=
\left(\ell\sigma_3\right)^{-n} \ell_n(u) \left(\ell\sigma_3\right)^n \mathbb{T}_{n-1}(u) \\
\nonumber
&= \left(\begin{array}{cc}
\frac{u}{\ell}+\frac{1}{\ell} a^{\dagger}_n a_n  &
\frac{(-1)^n}{\ell} a_n \\
2(-1)^na^{\dagger}_n+ \frac{(-1)^n}{\ell} a^{\dagger 2}_n a_n &
- \frac{u}{\ell}+\frac{1}{\ell} a^{\dagger}_n a_n
\end{array} \right) \mathbb{T}_{n-1}(u).
\end{align}
Now, we define renormalized creation and annihilation
operators
\begin{align}\nonumber
& \psi_n = (-1)^n\frac{b}{2} a_n; \ \bar{\psi}_n =
\ell (-1)^n a a^{\dagger}_n; \\
\label{com}
& \left[\psi_n,\bar{\psi}_m\right] =
\ell \frac{ab}{2} \bigl[a_n, a^{\dagger}_m\bigl] =
\ell \frac{ab}{2} \delta_{n m},
\end{align}
so that we could interpret the commutation relations~\eqref{com}
as a discrete version of the canonical commutation relations
for the fields. Let $\Delta$ be the lattice spacing; then
$x = n\Delta$, $y = m\Delta$ become continuous variables
as $\Delta \to 0$, and
\begin{align*}
\psi_{n} \to \psi(x); \quad
\bar{\psi}_{n} \to \bar{\psi}(x);
\quad \left[\psi_{n},\bar{\psi}_{m}\right] =
\frac{1}{\Delta} \delta_{n m} \to \left[\bar{\psi}(x),\psi(y)\right] = \delta(x-y).
\end{align*}
Taking into account~\eqref{com}, we get $\Delta = \frac{2}{ab}\,\frac{1}{\ell}$.
The difference on the left-hand side of expression~\eqref{diskr}
turns into the derivative $\mathbb{T}_n(u)-\mathbb{T}_{n-1}(u) \to
\Delta \partial_x \mathbb{T}(x, u)$ so that
\begin{align*}
\Delta \partial_x \mathbb{T}(x, u) =
\left(\begin{array}{cc}
\frac{u}{\ell}+
\frac{\Delta}{\ell} \bar{\psi}(x) \psi(x)  &
\frac{2}{b}\frac{1}{\ell} \psi(x) \\
\frac{2}{a}\frac{1}{\ell} \bar{\psi}(x)+ a \frac{\Delta}{\ell}
\bar{\psi}^2(x) \psi(x) &
- \frac{u}{\ell}- \frac{\Delta}{\ell} \bar{\psi}(x) \psi(x)
\end{array} \right)
\mathbb{T}(x, u).
\end{align*}
In the $\ell \to \infty$ limit we obtain
\begin{align*}
\partial_x \mathbb{T}(x, u) =
\left(\begin{array}{cc}
\frac{ab}{2} u  & a \psi(x) \\
b \bar{\psi}(x) & -\frac{ab}{2} u
\end{array} \right)
\mathbb{T}(x, u).
\end{align*}
This equation for the monodromy matrix coincides
with~\eqref{T0} after the renormalization of the
spectral parameter $u = \frac{1}{ab}\lambda$.

\subsection{Local Hamiltonian in the continuum limit}

To find a natural interpretation of the emerged
parameter $ab$, we shall take the $\ell \to \infty$
limit in the local Hamiltonian
$H = \sum_{k=1}^{n} H_{k+1 k}$~\cite{TTF}
\begin{align*}
H_{k+1 k} & = \psi\left(a^{\dagger}_k a_k -a^{\dagger}_{k+1} a_k+2\ell\right)+
\psi\left(a^{\dagger}_{k+1} a_{k+1} - a^{\dagger}_{k} a_{k+1}+2\ell\right)
- 2\psi(2\ell) \\
& = \psi\left(\Delta \bar{\psi}_k \psi_k + \Delta \bar\psi_{k+1} \psi_k+2\ell\right)+
\psi\left(\Delta \bar{\psi}_{k+1} \psi_{k+1} +
\Delta \bar\psi_{k} \psi_{k+1} +2\ell\right)
- 2\psi(2\ell).
\end{align*}
First recall that $x = k\Delta$ becomes a continuous variable
as $\Delta \to 0$; secondly,
$\psi_{k+1} \to \psi(x)+\Delta\partial\psi(x)+\frac{\Delta^2}{2}\partial^2\psi(x)+\ldots$
and the same for $\bar{\psi}_{k+1}$. Using these formulas,
the asymptotic expansion of the logarithmic derivative
of the gamma function
$$
\psi(z) \to \ln z -\frac{1}{2z}-\frac{1}{12z^2}+\ldots
$$
and replacing sums by integrals $\Delta\sum_k \to \int dx$,
we get~\cite{TTF}
\begin{align*}
H & = \sum_{k=1}^{n} H_{k+1 k}  \\
\to & \left(\frac{2}{\ell}+\frac{1}{2\ell^2}+\frac{1}{12\ell^3}\right)
\int d x\, \bar{\psi}(x)\psi(x) -\Delta^2\frac{1}{2\ell}
\int d x\,\left(\partial\bar{\psi}(x)\partial\psi(x)+
ab \,\bar{\psi}^2(x)\psi^2(x)\right)
+\ldots,
\end{align*}
where we have written all terms to order $\frac{1}{\ell^4}$.
Terms with total derivatives $\partial\left(\bar{\psi}(x)\psi(x)\right)$
are canceled due to periodic boundary conditions. 
Instead of the original local Hamiltonian we can consider the operator
($\Delta = \frac{2}{ab} \frac{1}{\ell}$)
\begin{align*}
H^{\prime} = \sum_{k=1}^{n}\left( H_{k+1 k} - \Delta \left(\frac{2}{\ell}+\frac{1}{2\ell^2}+
\frac{1}{12\ell^3}\right)\bar{\psi}_k\psi_k\right),
\end{align*}
which also commutes with the transfer matrix and describes
the interaction between the nearest neighbors. By the same
argument, we get the expansion
\begin{align*}
H^{\prime} \to  -\Delta^2\frac{1}{2\ell}
\int d x\,\left(\partial\bar{\psi}(x)\partial\psi(x)+
ab\, \bar{\psi}^2(x)\psi^2(x)\right)
+\ldots
\end{align*}
Thus in the continuum limit we obtain the standard
Hamiltonian of the continuous model~\cite{S1,F3,KBI}
\begin{align*}
H = \int d x\,\left(\partial\bar{\psi}(x)\partial\psi(x)+
ab\, \bar{\psi}^2(x)\psi^2(x)\right),
\end{align*}
and $ab = c$ is the coupling constant.

\subsection{$\mathrm{Q}$-operator in the continuum limit}

We know the explicit connection between the discrete and continuous 
models so that it is possible to calculate the $\ell \to \infty$ 
limit in the $\mathrm{Q}$-operator
\begin{equation} \label{Q-spin}
Q(u) = \frac{\Gamma^n(2 \ell)}{\Gamma^n(\ell + u)} R_{n n-1} (u) \cdots R_{10}(u)
\bigl|_{z_0 \rightarrow z_n}
\end{equation}
expressed in terms of the discrete fields $\psi_k$, $\bar\psi_k$.
Before we plunge into the calculation, recall that the eigenvalues
of the operator $Q(u)$ depend on spin $\ell$
$$
Q(u) |v_1 \ldots v_l \rangle = \frac{u - v_1}{\ell - v_1} \cdots \frac{u -
v_l}{\ell - v_l} |v_1 \ldots v_l \rangle.
$$
So, before taking the limit we should normalize the
$\mathrm{Q}$-operator appropriately. One way of
doing it
$$
( \ell/u )^{\sum z_k \pd_k} Q(u) |v_1 \ldots v_l \rangle = \prod \frac{\ell}{u}
\frac{u - v_k}{ \ell - v_k} \, |v_1 \ldots v_l \rangle.
$$
Here and further we mean the summation over all $n$ sites of chain.
The operator $\sum z_k \pd_k$ counts the number of $v_i$-parameters in
a given state $|v_1 \ldots v_l \rangle$, and we put the parameter $u$ in the denominator for the sake of convenience of the final result.

To take the continuum limit we use the integral form of the
$\mathrm{R}$-operators
$$
R_{k k-1}(u) = \frac{1}{\Gamma(\ell - u)} \int_0^1 d\alpha_k \, \alpha_k^{u +
\ell - 1} (1 - \alpha_k)^{\ell - u - 1} \alpha_k^{z_{k k-1} \pd_k}.
$$
The last factor in this expression acts nontrivially on the variable $z_k$ only
$$
\alpha_k^{z_{k k-1} \pd_k} \, \Phi(z_k) = \Phi (\alpha_k z_k + (1 - \alpha_k) z_{k-1} ).
$$
Together with the normalization part from~\eqref{Q-spin} the product of these operators
acts on functions of all variables $z_k$ ($k = 1, \ldots, n$)
$$
(\ell / u)^{\sum z_k \pd_k} \left( \prod \alpha_k^{z_{k k-1} \pd_k} \right) \! \Bigl|_{z_0 \rightarrow z_n}
 \, \Phi (\ldots, z_k, \ldots) = \Phi (\ldots, (\ell/u) \alpha_k z_k + (\ell/u) (1 - \alpha_k) z_{k-1}, \ldots).
$$
It is useful to rewrite this operator in a normal ordered form.
Consider the following operator
$$
\normord{e^{(c_1 z_k + c_2 z_{k-1}) \pd_k}} = \sum_{p=0}^\infty \frac{1}{p!}
(c_1 z_k + c_2 z_{k-1})^p \pd_k^p.
$$
Under the sign of the normal ordering  the operators
$z_k$ and $\partial_k$ commute. Acting on the monomial
\begin{equation*}
\normord{e^{(c_1 z_k + c_2 z_{k-1}) \pd_k}} \, z_k^p = z_k^p + p (c_1 z_k
+ c_2 z_{k-1}) z_k^{p-1}
+ \ldots + (c_1 z_k + c_2 z_{k-1})^p = ((c_1 + 1) z_k + c_2 z_{k-1})^p,
\end{equation*}
for an arbitrary function of $z_k$ we get
$$
\normord{e^{(c_1 z_k + c_2 z_{k-1}) \pd_k}} \, \Phi(z_k) = \Phi((c_1 + 1) z_k +
c_2 z_{k-1}).
$$
Note also that for the neighboring operators
$$
\normord{e^{(c_1 z_{k+1} + c_2 z_k) \pd_{k+1}}}
\, \normord{ e^{(c_3 z_k + c_4 z_{k-1}) \pd_k}}
= \normord{e^{(c_1 z_{k+1} + c_2 z_k) \pd_{k+1}}
e^{(c_3 z_k + c_4 z_{k-1}) \pd_k}}
$$
since $z_k$ and $z_{k-1}$ commute with $\pd_{k+1}$.
By taking appropriate coefficients $c_{i}$, we find
the normal ordered form of the original operators
$$
(\ell / u)^{\sum z_k \pd_k} \left( \prod \alpha_k^{z_{k k-1} \pd_k} \right)
\! \Bigl|_{z_0 \rightarrow z_n} = \normord{\prod e^{(\ell/u - 1)z_k \pd_k
+ (\ell/u)(1 - \alpha_k) z_{k-1 k} \pd_k}}
$$

Thus we have the following formula for the renormalized
$\mathrm{Q}$-operator
$$
(\ell/u)^{\sum z_k \pd_k} Q(u) = \normord{G(u)}
$$
where
$$
G(u) = \left[ \frac{\Gamma \left( 2 \ell \right) }{\Gamma
\left( \ell + u \right) \Gamma \left( \ell - u \right)} \right]^n \prod_{k=1}^n
\int_0^1 d\alpha_k \, \alpha_k^{u+ \ell - 1} (1 - \alpha_k)^{\ell - u -1}
e^{(\ell/u - 1)a^{\dagger}_k a_k + (\ell/u)(1 - \alpha_k)
\bigl(a^{\dagger}_{k-1}-a^{\dagger}_{k}\bigl) a_k}.
$$
Under the sign of the normal ordering the operators 
$a^{\dagger}_k = z_k$ and $a_k = \partial_k$ commute, hence in what follows all fields are treated as classical.
The continuous model is defined on the finite interval. 
Denote by $x$ and $y$ the upper and lower endpoints of
this interval so that $(x-y) = n \Delta$.  Since the length
of the interval in the $\ell \to \infty$ limit remains constant,
the number of chain sites $n \sim \ell$. In other words,
the number of integrals tends to infinity. Below we will
show that, in fact, in the continuum limit this multiple integral
boils down to some functional integral.

Let us change integration variables in each integral
$$
\alpha_k = \frac{1}{2}\left( 1 + \frac{\beta_k}{\ell} \right), \qquad
1 - \alpha_k = \frac{1}{2}\left( 1 - \frac{\beta_k}{\ell} \right)
$$
and pass from the original operators $a^{\dagger}_k$ and $a_k$
to the renormalized ones
\begin{gather*}
a^{\dagger}_k a_k = \Delta \bar\psi_k \psi_k, \\
\left(a^{\dagger}_{k-1}-a^{\dagger}_{k}\right) a_k =
\Delta
\left(\bar\psi_{k}-\bar\psi_{k-1}\right) \psi_k - 2 \Delta \bar\psi_k \psi_k.
\end{gather*}
Then we have
\begin{multline*}
G(u) = \mathcal{N} \int_{-\ell}^\ell
d\beta_1 \ldots \int_{-\ell}^\ell d\beta_n \, \exp \Biggl[ (\ell - 1) \sum \ln \left(1 - \frac{\beta_k^2}{\ell^2}
\right) + u \sum \ln \frac{1 + \nicefrac{\beta_k}{\ell}}{1 - \nicefrac{\beta_k}{\ell}} \\
+ \Delta \sum \left( \frac{\beta_k}{u} - 1 \right) \bar\psi_k \psi_k +
\frac{\ell}{2u}\Delta \sum \left(1 - \frac{\beta_k}{\ell}\right)
\left(\bar\psi_{k}-\bar\psi_{k-1}\right) \psi_k \Biggl],
\end{multline*}
where the factor in front of the integral equals
$$
\mathcal{N} = \left[ \frac{2}{4^\ell \, \ell \, \mathrm{B}(\ell+u,\ell-u)} \right]^n.
$$
Using Stirling's formula for the beta function, it's easy to calculate
its asymptotic behaviour
$$
\mathcal{N} \sim \left[ \frac{1}{\sqrt{\pi \ell}} \right]^n \exp \left[ - \frac{u^2(x-y)c}{2} \right].
$$
The sums in the exponent can be interpreted as some discretized
functional of the original fields $\psi(t)$ and $\bar\psi(t)$
and one auxiliary field $\beta(t)$. In the $\ell \to \infty$ limit
each sum turns into the corresponding integral. For example, for the
first sum we have
$$
(\ell - 1) \sum \ln \left( 1 - \frac{\beta_k^2}{\ell^2} \right) = - (\ell - 1) \sum \frac{\beta_k^2}
{\ell^2} + \cdots \longrightarrow - \frac{c}{2} \int_y^x dt \, \beta^2(t),
$$
where we have used the fact that $\frac{1}{\ell} = \frac{c\Delta}{2}$. 
For the remaining sums everything is similar so that
in the continuum limit we obtain 
\begin{multline*}
G(u) = \int \mathcal{D}\beta \, \exp \Biggl[ - \frac{u^2(x-y)c}{2} - \frac{c}{2} \int_y^x dt \, \beta^2(t) + c u \int_y^x dt \, \beta(t)
\\
+ \int_y^x dt \left( \frac{1}{u} \beta(t) - 1 \right) \bar\psi(t) \psi(t) + \frac{1}{cu} \int_y^x dt \, \pd_t \bar\psi(t) \psi(t) \Biggl].
\end{multline*}
Here we introduced the notation for the integration measure
$$
\mathcal{D} \beta = \lim_{\ell \rightarrow \infty} \frac{d\beta_1}{\sqrt{\pi \ell}} \cdots \frac{d\beta_n}{\sqrt{\pi \ell}}.
$$
In order to check the norm, one could consider a simpler integral:
$$
\int \mathcal{D} \beta \exp \left[ - \frac{c}{2} \int_y^x dt \, \beta^2(t) \right] =
\lim_{\ell \rightarrow \infty} \prod_k \int \frac{d\beta_k}{\sqrt{\pi \ell}}
\exp \left[ - \frac{1}{\ell} \beta_k^2 \right] = 1.
$$
Note that we can simplify the obtained formula by changing integration
variables
\begin{multline*}
G(u) = \int \mathcal{D}\beta \, \exp \Biggl[ - \frac{c}{2} \int_y^x dt \, (\beta(t) - u)^2
+ \frac{1}{u} \int_y^x dt \left( \beta(t) - u \right) \bar\psi(t) \psi(t) + \frac{1}{cu} \int_y^x dt \, \pd_t \bar\psi(t) \psi(t) \Biggl] \\
= \int \mathcal{D} \phi \exp \left[ -\frac{c}{2} \int_y^x dt \, \phi^2(t) + \frac{1}{u} \int_y^x dt \, \bar\psi(t) \psi(t) \phi(t) + \frac{1}{cu} \int_y^x dt \, \pd_t \bar\psi(t) \psi(t) \right].
\end{multline*}
Moreover, this functional integral is Gaussian, so we can compute it
$$
G(u) = \exp \left[ -\frac{1}{cu} \int_y^x dt \, \bar\psi(t) \pd_t \psi(t)
+ \frac{1}{2cu^2} \int_y^x dt \, (\bar\psi(t) \psi(t))^2 \right].
$$
Additional terms appearing after integration by parts in the first
term cancel each other due to periodic boundary conditions.

Recall that in the continuum limit we obtained the following
equation for the monodromy matrix
\begin{align*}
\partial_x \mathbb{T}(x,u) =
\left(\begin{array}{cc}
\frac{ab}{2} u  & a\psi(x) \\
b \bar{\psi}(x) & -\frac{ab}{2} u
\end{array} \right)
\mathbb{T}(x,u),
\end{align*}
which coincides with~\eqref{T0} after the renormalization of the
spectral parameter $u = \frac{1}{ab}\lambda$. In these terms,
the $\mathrm{Q}$-operator for the continuous model has the
following form
\begin{align}\label{QQQ}
\mathbb{Q}^x_y(\lambda) = \lim_{\ell \rightarrow \infty} (c \ell/\lambda)^{\sum z_k \pd_k} Q(\lambda/c) =
\normord{\exp \left[ -\frac{1}{\lambda} \int_y^x dt \, \bar\psi(t) \pd_t \psi(t)
+ \frac{c}{2\lambda^2} \int_y^x dt \, (\bar\psi(t) \psi(t))^2 \right]}
\end{align}

\section{Continuous model}

\subsection{Monodromy matrix}

The classical monodromy matrix $T^{x}_{y}(\lambda)$ is defined by
the differential equation~\cite{S1,F3}
\begin{align}\label{T0}
&\partial_x T^{x}_{y}(\lambda) = L(x,\lambda)T^{x}_{y}(\lambda),
\quad T^{x}_{x}(\lambda) =
\left( \begin{array}{cc}
1 & 0 \\
0 & 1 \\
\end{array} \right) \\
\nonumber
& L(x,\lambda) = \frac{\lambda}{2} \sigma_3+a\psi(x) \sigma_{+}+
b\bar{\psi}(x) \sigma_{-} =
\left( \begin{array}{cc}
\nicefrac{\lambda}{2} & a \psi(x) \\
b \bar{\psi}(x) & -\nicefrac{\lambda}{2} \\
\end{array} \right),
\end{align}
which can be solved using the ordered exponential
\begin{align}\label{T}
T^{x}_{y}(\lambda) = \overset{\curvearrowleft}{\exp}
\int_{y}^{x} L(t,\lambda) d t =
\II + \int_{y}^{x} d t\, L(t,\lambda) +
\int_{y}^{x}d t_1\, L(t_1,\lambda)
\int_{y}^{t_1}d t_2\, L(t_2,\lambda) + \ldots
\end{align}
The entries of the monodromy matrix
$$
T^{x}_{y}(\lambda) = \left( \begin{array}{cc}
A^{x}_{y}(\lambda) & B^{x}_{y}(\lambda) \\
C^{x}_{y}(\lambda) & D^{x}_{y}(\lambda) \\
\end{array} \right)
$$
are functionals of the fields $\psi$, $\bar{\psi}$, hence
the full notation for the monodromy matrix should be
$T^{x}_{y}(\psi, \bar{\psi}, \lambda)$. To make formulas
more readable, we omit the dependence on $\lambda$
in the expressions containing functional derivatives with
respect to the fields, and similarly, we drop the dependence
on the fields in the equations with no functional derivatives.

In the quantum NLS model~\cite{S1,F1,F3,KBI,N} the
fields obey the canonical commutation relations
\begin{align*}
\left[\bar{\psi}(x),\bar{\psi}(y)\right] = \left[\psi(x),\psi(y)\right] = 0;
\quad \left[\psi(x),\bar{\psi}(y)\right] = \delta(x-y).
\end{align*}
The global quantum space is a Fock space $\mathbb{F}$,
and the vacuum vector is defined in a standard way
\begin{align*}
\psi(x) |0\rangle = 0.
\end{align*}
The Hamiltonian of the model is given by
\begin{align*}
H = \int d x\,\left(\partial\bar{\psi}(x)\partial\psi(x)+
c\,\bar{\psi}^2(x)\psi^2(x)\right),
\end{align*}
where $c = ab$ is the coupling constant.

The quantum monodromy matrix $\mathbb{T}^{x}_{y}(\lambda)$
is defined as~\cite{S1,F1,KBI}
\begin{align}\label{Tq}
\mathbb{T}^{x}_{y}(\lambda) = \left( \begin{array}{cc}
\mathbb{A}^{x}_{y}(\lambda) & \mathbb{B}^{x}_{y}(\lambda) \\
\mathbb{C}^{x}_{y}(\lambda) & \mathbb{D}^{x}_{y}(\lambda) \\
\end{array} \right) = \normord{T^{x}_{y}(\lambda)} = \left( \begin{array}{cc}
\normord{A^{x}_{y}(\lambda)} & \normord{B^{x}_{y}(\lambda)} \\
\normord{C^{x}_{y}(\lambda)} & \normord{D^{x}_{y}(\lambda)} \\
\end{array} \right).
\end{align}
In other words, a normal (or Wick) symbol of the quantum monodromy
matrix coincides with the classical monodromy matrix.
The quantum monodromy matrix acts in a tensor product of two spaces:
one auxiliary space $\mathbb{C}^2$ and the Fock space $\mathbb{F}$,
$$
\mathbb{T}^{x}_{y}(\lambda) \colon \mathbb{C}^2\otimes \mathbb{F} \to \mathbb{C}^2\otimes\mathbb{F}.
$$
This implies that the entries of the quantum monodromy matrix
are the operators acting in the Fock space, and
$\mathbb{T}^{x}_{y}(\lambda)$ as a matrix acts nontrivially
in the auxiliary space $\mathbb{C}^2$.

Commutation relations between the entries of the
quantum monodromy matrices $\mathbb{T}^{x}_{y}(\lambda)$
and $\mathbb{T}^{x}_{y}(\mu)$ can be written compactly as~\cite{S1,F1}
\begin{equation}\label{glob}
\mathbb{R}_{12}(\lambda-\mu)\,
\mathbb{T}_1(\lambda) \mathbb{T}_2(\mu) =
\mathbb{T}_2(\mu) \mathbb{T}_1(\lambda)
\mathbb{R}_{12}(\lambda-\mu).
\end{equation}
Here all operators act in a tensor product of three spaces
$\mathbb{V}_1\otimes \mathbb{V}_2\otimes\mathbb{F}$,
where $\mathbb{V}_1 = \mathbb{C}^2$ and $\mathbb{V}_2 = \mathbb{C}^2$
are two auxiliary spaces. The operator
$\mathbb{T}_1(\lambda) = \mathbb{T}^{x}_{y}(\lambda)\otimes\II$
as a matrix acts nontrivially in the first space $\mathbb{V}_1$.
Its matrix elements act in the Fock space. In the second auxiliary
space $\mathbb{V}_2$ the operator $\mathbb{T}_1(\lambda)$
acts as an identity matrix. In the same way, the operator
$\mathbb{T}_2(\mu) = \II\otimes\mathbb{T}^{x}_{y}(\mu)$
acts nontrivially in the $\mathbb{V}_2$ and $\mathbb{F}$.
Finally the $\mathrm{R}$-matrix is defined as
\begin{align}\label{R}
\mathbb{R}_{12}(\lambda-\mu) = (\lambda-\mu)\II - c \mathbb{P}_{12},
\end{align}
where by $\mathbb{P}_{12}$ we denote a permutation operator~\eqref{perm}.
The proof of the commutation relations~\eqref{glob} is given in
Appendix~\ref{C}.

\subsection{$\mathrm{Q}$-operator}

\subsubsection{Commutation relations with the monodromy matrix}

Consider the operator~\eqref{QQQ}
\begin{align}\label{QZ}
& \mathbb{Q}^{x}_{y}(\lambda) =
\normord{e^{S(\bar\psi,\psi,\lambda)}} =
\normord{e^{-\frac{1}{\lambda}\bar{\psi}\partial\psi +
\frac{ab}{2\lambda^2}\left(\bar{\psi}\psi\right)^2}} \\
\nonumber
& S(\bar\psi,\psi,\lambda) = -\frac{1}{\lambda}
\int_{y}^{x} \bar{\psi}(t)\partial_t\psi(t) dt +
\frac{ab}{2\lambda^2}
\int_{y}^{x} \left(\bar{\psi}(t)\psi(t)\right)^2 dt.
\end{align}
Below we prove that this operator satisfies all
the needed properties of a $\mathrm{Q}$-operator.

In what follows we use the so-called \textit{universal
notations}~\cite{V} and omit the obvious integral
symbol, the arguments of the fields, etc. The example is
given in formula~\eqref{QZ}, where the exponent
$S(\bar\psi,\psi,\lambda) = -\frac{1}{\lambda}\bar{\psi}\partial\psi +\frac{ab}{2\lambda^2}\left(\bar{\psi}\psi\right)^2$
is written using universal notation, and below it is written
in a full form.

So, first let us prove that the operator $\mathbb{Q}^{x}_{y}(\lambda)$
commutes with the transfer matrix. To do this, we want to
rewrite the product of the $\mathrm{Q}$-operator and
the monodromy matrix in the normal ordered form.
Representing the quantum monodromy
matrix~\eqref{Tq} as
\begin{align*}
\mathbb{T}^{x}_{y}(\mu) = \normord{T^{x}_{y}(\psi,\bar{\psi},\mu)} =
\left.T^{x}_{y}\left(\frac{\delta}{\delta \bar{A}},
\frac{\delta}{\delta A},\mu\right)
\normord{e^{A\bar{\psi}}\, e^{\bar{A}\psi}}
\right|_{A=\bar{A}=0}
\end{align*}
and using identity~\eqref{a}
\begin{align*}
\mathbb{Q}^{x}_{y}(\lambda) e^{A\bar{\psi}} e^{\bar{A}\psi} =
\normord{\exp \left( S(\bar\psi,\psi+A,\lambda) + A\bar{\psi} +\bar{A}\psi\right)}
\end{align*}
we get
\begin{align*}
\mathbb{Q}^{x}_{y}(\lambda)
\mathbb{T}^{x}_{y}(\mu) & =
\left.T^{x}_{y}\left(\frac{\delta}{\delta \bar{A}},
\frac{\delta}{\delta A},\mu\right)\normord{ e^{S(\bar\psi,\psi,\lambda) +
A\left(\bar{\psi}+\frac{1}{\lambda}\partial\bar{\psi} +
\frac{ab}{\lambda^2}\bar{\psi}^2\psi\right) +\bar{A}\psi+
\underline{\frac{ab}{\lambda^2}\left(\bar{\psi}A\right)^2}}}
\right|_{A=\bar{A}=0} \\
& = \left.T^{x}_{y}\left(\frac{\delta}{\delta \bar{A}},
\frac{\delta}{\delta A},\mu\right)\normord{ e^{S(\bar\psi,\psi,\lambda) +
A\left(\bar{\psi}+\frac{1}{\lambda}\partial\bar{\psi} +
\frac{ab}{\lambda^2}\bar{\psi}^2\psi\right) +\bar{A}\psi}}
\right|_{A=\bar{A}=0}.
\end{align*}
Here we use the universal notation again. As an example,
we write two terms in the full form
$$
\bar{A}\psi = \int_{y}^{x} \bar{A}(t)\psi(t) dt, \quad
\left(\bar{\psi}A\right)^2 = \int_{y}^{x} A^2(t)\bar\psi^2(t) dt
$$
and hope that further the notation will be clear. Let us remark that
the underlined term $\frac{ab}{\lambda^2}\left(\bar{\psi}A\right)^2$
in the exponent leads to the additional local terms
corresponding to higher functional derivatives, for instance,
$$
\frac{\delta^2}{\delta A(t_1)\delta A(t_2)} \to
\ldots + \frac{2ab}{\lambda^2}\bar{\psi}^2(t_1)\delta(t_1-t_2).
$$
In expression for the monodromy matrix~\eqref{T}
the strict order of parameters $t_i$ (such as $t_1 > t_2$, etc.)
is required. Hence the local terms don't contribute to the result.

Calculating the functional derivatives
\begin{align*}
\frac{\delta}{\delta A(t)} \to
\bar{\psi}(t)+\frac{1}{\lambda}\partial_t\bar{\psi}(t) +
\frac{ab}{\lambda^2}\bar{\psi}^2(t)\psi(t); \quad
\frac{\delta}{\delta \bar{A}(t)} \to \psi(t),
\end{align*}
we obtain
\begin{align*}
\mathbb{Q}^{x}_{y}(\lambda)
\mathbb{T}^{x}_{y}(\mu) =
\normord{ e^{S(\bar\psi,\psi,\lambda)}
T^{x}_{y}\left(\psi,\bar{\psi}+\frac{1}{\lambda}\partial\bar{\psi} +
\frac{ab}{\lambda^2}\bar{\psi}^2\psi,\mu\right)} = \normord{ e^{S(\bar\psi,\psi,\lambda)}
T^{x}_{y}\left(\lambda,\mu\right)}
\end{align*}
where the monodromy matrix under the sign of the normal ordering 
is defined by the differential equation
\begin{align}\nonumber
& \partial_xT^{x}_{y}(\lambda,\mu) = L(x,\lambda,\mu)T^{x}_{y}(\lambda,\mu),
\quad T^{x}_{x}(\lambda,\mu) = \II \\
\label{Lml}
& L(x,\lambda,\mu) = \frac{\mu}{2}\sigma_3+a\psi(x)\sigma_{+}+
b\left(\bar{\psi}(x)+\frac{1}{\lambda}\partial\bar{\psi}(x) +
\frac{ab}{\lambda^2}\bar{\psi}^2(x)\psi(x)\right) \sigma_{-}.
\end{align}
It is easy to check that the matrix $L(x,\lambda,\mu)$ can be transformed 
to the much simpler matrix $U(x,\lambda,\mu)$ by the appropriate gauge transformation
\begin{align*}
& L(x,\lambda,\mu) = S^{-1}(x,\lambda)
U(x,\lambda,\mu)\,S(x,\lambda) -
S^{-1}(x,\lambda)\partial_x S(x,\lambda) \\
& U(x,\lambda,\mu) = \left(\frac{\mu}{2}+\frac{ab}{\lambda}\bar{\psi}\psi
\right)\sigma_3+b\left(1-\frac{\mu}{\lambda}\right)\bar{\psi}\sigma_{-}+
a\psi \sigma_{+}; \quad S(x,\lambda) = \II-\frac{b}{\lambda}\bar{\psi}(x)\sigma_{-}.
\end{align*}
Then
\begin{align}\label{S}
T^{x}_{y}(\lambda,\mu) = S^{-1}(x,\lambda) F^{x}_{y}(\lambda,\mu) S(y,\lambda),
\end{align}
where $F^{x}_{y}(\lambda,\mu)$ is a solution of the equation
\begin{align*}
& \partial_x F^{x}_{y}(\lambda,\mu) = U(x,\lambda,\mu) F^{x}_{y}(\lambda,\mu),
\quad F^{x}_{x}(\lambda,\mu) = \II \\
& U(x,\lambda,\mu) =
\left( \begin{array}{cc}
\nicefrac{\mu}{2}+\nicefrac{ab}{\lambda} \bar{\psi}(x)\psi(x) & a \psi(x) \\
b \left(1-\nicefrac{\mu}{\lambda}\right)\bar{\psi}(x) & -\nicefrac{\mu}{2}-\nicefrac{ab}{\lambda} \bar{\psi}(x)\psi(x)
\end{array} \right).
\end{align*}
In the case of periodic boundary conditions:
$\psi(x)=\psi(y)$, $\bar{\psi}(x)=\bar{\psi}(y)$. Hence
if we take the trace of both sides in formula~\eqref{S},
the matrices $S^{-1}(x,\lambda)$ and $S(y,\lambda)$
cancel each other, and for the product of the
$\mathrm{Q}$-operator and the transfer matrix we get
\begin{align*}
\mathbb{Q}^{x}_{y}(\lambda)\,
\tr \mathbb{T}^{x}_{y}(\mu) = \normord{ e^{S(\bar\psi,\psi,\lambda)}\,
\tr F^{x}_{y}\left(\lambda,\mu\right)}
\end{align*}
Now consider the product of the $\mathrm{Q}$-operator
and the transfer matrix in the reverse order. As above,
we use the identity~\eqref{a} and obtain a shift of
the field $\bar{\psi} \to \bar{\psi} + \bar{A}$
\begin{align*}
e^{A\bar{\psi}} e^{\bar{A}\psi} \mathbb{Q}^{x}_{y}(\lambda) &=
\normord{\exp \left( S(\bar\psi+\bar{A},\psi,\lambda) + A\bar{\psi} +\bar{A}\psi\right)} \\
& = \normord{\exp \left( S(\bar\psi,\psi,\lambda) +
\bar{A}\left(\psi-\frac{1}{\lambda}\partial \psi +
\frac{ab}{\lambda^2}\psi^2\bar{\psi}\right) +\bar{A}\psi+
\frac{ab}{\lambda^2}\left(\psi\bar{A}\right)^2\right)}
\end{align*}
By differentiating with respect to the sources
\begin{align*}
\frac{\delta}{\delta \bar{A}(t)} \to \psi(t)-\frac{1}{\lambda}\partial_t\psi(t) +
\frac{ab}{\lambda^2}\psi^2(t)\bar{\psi}(t); \quad
\frac{\delta}{\delta A(t)} \to \bar{\psi}(t),
\end{align*}
we find
\begin{align*}
\mathbb{T}^{x}_{y}(\mu)
\mathbb{Q}^{x}_{y}(\lambda) =
\normord{ e^{S(\bar\psi,\psi,\lambda)}
T^{\prime x}_{y}\left(\mu,\lambda\right)}
\end{align*}
where the matrix $T^{\prime x}_{y}\left(\mu,\lambda\right)$
is a solution of the equation
\begin{align}\nonumber
& \partial_x T^{\prime x}_{y}\left(\mu,\lambda\right) = L^{\prime}(x,\mu,\lambda)
T^{\prime x}_{y}\left(\mu,\lambda\right), \quad
T^{\prime x}_{x}\left(\mu,\lambda\right) = \II \\
\label{Llm}
& L^{\prime}(x,\mu,\lambda) = \frac{\mu}{2} \sigma_3+
a\left(\psi(x)-\frac{1}{\lambda}\partial\psi(x) +
\frac{ab}{\lambda^2}\psi^2(x)\bar{\psi}(x)\right) \sigma_{+}+
b\bar{\psi}(x) \sigma_{-}.
\end{align}
We use the gauge transformation again
\begin{align*}
& L^{\prime}(x,\mu,\lambda) = S^{\prime -1}(x,\lambda)
U^{\prime}(x,\mu,\lambda) S^{\prime}(x,\lambda) -
S^{\prime -1}(x ,\lambda)\partial_x S^{\prime}(x ,\lambda) \\
& U^{\prime}(x,\mu,\lambda) = \left(\frac{\mu}{2}+\frac{ab}{\lambda}\bar{\psi}\psi
\right) \sigma_3+b\bar{\psi}\sigma_{-}+
a\left(1-\frac{\mu}{\lambda}\right)\psi \sigma_{+};  \quad S^{\prime}(x,\lambda) =
\II+\frac{a}{\lambda}\psi(x)\sigma_{+}\,,
\end{align*}
so that 
\begin{align}\label{S'}
T^{\prime x}_{y}\left(\mu,\lambda\right) = S^{\prime -1}(x ,\lambda) G^{x}_{y}\left(\mu, \lambda\right)
S^{\prime}(y,\lambda),
\end{align}
where $G^{x}_{y}(\mu, \lambda)$ satisfies the differential equation
\begin{align*}
& \partial_x G^{x}_{y}(\mu,\lambda) = U^{\prime}(x ,\mu ,\lambda) G^{x}_{y}(\mu ,\lambda),
\quad G^{x}_{x}(\mu ,\lambda) = \II \\
& U^{\prime}(x ,\mu ,\lambda) =
\left( \begin{array}{cc}
\nicefrac{\mu}{2}+\nicefrac{ab}{\lambda} \bar{\psi}(x)\psi(x) & a \left(1-\nicefrac{\mu}{\lambda}\right)\psi(x) \\
b \bar{\psi}(x) & -\nicefrac{\mu}{2}-\nicefrac{ab}{\lambda}
\bar{\psi}(x)\psi(x)
\end{array} \right).
\end{align*}
As before, the matrices $S^{\prime  -1}(x, \lambda)$ and
$S^{\prime}(y ,\lambda)$ in~\eqref{S'} cancel each other within
the trace due to periodic boundary conditions. Thus for the product
of the $\mathrm{Q}$-operator and the transfer matrix in the reverse
order we get
\begin{align*}
\tr \mathbb{T}^{x}_{y}(\mu)\,
\mathbb{Q}^{x}_{y}(\lambda) = \normord{ e^{S(\bar\psi,\psi,\lambda)}\,
\tr\, G^{x}_{y}\left(\mu,\lambda\right)}
\end{align*}
Notice that the matrices $U^{\prime}(x,\mu,\lambda)$ and
$U(x,\lambda,\mu)$ are connected by the similarity transformation
\begin{align*}
\left( \begin{array}{cc}
1 & 0 \\
0 & 1-\nicefrac{\mu}{\lambda}
\end{array} \right) U^{\prime}(x,\mu,\lambda) =
U(x,\lambda,\mu)\left( \begin{array}{cc}
1 & 0 \\
0 & 1-\nicefrac{\mu}{\lambda}
\end{array} \right),
\end{align*}
hence a similar formula holds for the matrices
$G^{x}_{y}(\mu,\lambda)$ and $F^{x}_{y}(\lambda,\mu)$.
So we have
$\tr\, G^{x}_{y}(\mu,\lambda) = \tr F^{x}_{y}(\lambda,\mu)$.
This proves that the $\mathrm{Q}$-operator~\eqref{QZ} commutes
with the transfer matrix
\begin{align*}
\tr \mathbb{T}^{x}_{y}(\mu)\,
\mathbb{Q}^{x}_{y}(\lambda) = \mathbb{Q}^{x}_{y}(\lambda)\,
\tr \mathbb{T}^{x}_{y}(\mu).
\end{align*}

Now we derive the Baxter equation. Actually we already
have all the needed formulas. Consider the product of the
$\mathrm{Q}$-operator and the transfer matrix with
the same spectral parameter $\lambda$
\begin{align*}
\mathbb{Q}^{x}_{y}(\lambda)\,
\tr \mathbb{T}^{x}_{y}(\lambda) =
\normord{ e^{S(\bar\psi,\psi,\lambda)}\,
\tr F^{x}_{y}\left(\lambda\right)}
\end{align*}
where $F^{x}_{y}(\lambda)$ is a solution of the equation
\begin{align*}
& \partial_x F^{x}_{y}(\lambda) = U(x, \lambda) F^x_y(\lambda),
\quad F^{x}_{x}(\lambda) = \II \\
& U(x , \lambda) =
\left( \begin{array}{cc}
\nicefrac{\lambda}{2}+\nicefrac{ab}{\lambda} \bar{\psi}(x)\psi(x) & a \psi(x) \\
0 & -\nicefrac{\lambda}{2}-\nicefrac{ab}{\lambda} \bar{\psi}(x)\psi(x)
\end{array} \right).
\end{align*}
For the calculation of the trace, we need the diagonal elements
of the matrix $F^{x}_{y}(\lambda)$ only. Since the matrix $U(x,\lambda)$
is triangular, we find
\begin{align*}
F^{x}_{y}(\lambda) =
\left( \begin{array}{cc}
\exp \left(\frac{(x-y)\lambda}{2}+
\frac{ab}{\lambda}\int_y^x \bar{\psi}(t)\psi(t) d t\right) & \ldots \\
0 & \exp \left(-\frac{(x-y)\lambda}{2}-
\frac{ab}{\lambda}\int_y^x \bar{\psi}(t)\psi(t) d t\right)
\end{array} \right),
\end{align*}
and thus we derive the Baxter equation
\begin{align*}
& \mathbb{Q}^{x}_{y}(\lambda)\,
\tr  \mathbb{T}^{x}_{y}(\lambda) =
e^{\frac{(x-y)\lambda}{2}} \normord{ e^{S(\bar\psi ,\psi ,\lambda)+ \frac{ab}{\lambda} \bar{\psi}\psi}}+
e^{-\frac{(x-y)\lambda}{2}} \normord{ e^{S(\bar\psi ,\psi ,\lambda)- \frac{ab}{\lambda} \bar{\psi}\psi}} = \\
& = e^{\frac{(x-y)\lambda}{2}} \left(1+\frac{c}{\lambda}\right)^{\bar{\psi}\psi}
\mathbb{Q}^{x}_{y}(\lambda + c) +
e^{-\frac{(x-y)\lambda}{2}} \left(1-\frac{c}{\lambda}\right)^{\bar{\psi}\psi}
\mathbb{Q}^{x}_{y}(\lambda - c).
\end{align*}
Here we took into account that $ab = c$ and used formula~\eqref{b3}.

Finally we prove the commutation relation~\cite{Z}
\begin{align*}
\mathbb{Q}^{x}_{y}(\lambda)\,
\mathbb{Q}^{x}_{y}(\mu) =
\mathbb{Q}^{x}_{y}(\mu)\,
\mathbb{Q}^{x}_{y}(\lambda)
\end{align*}
Recall from the previous section that the $\mathrm{Q}$-operator
can be expressed as a functional integral
\begin{align*}
& \mathbb{Q}^{x}_{y}(\lambda) =
\normord{\int \mathcal{D} \phi \,
e^{S(\phi ,\bar\psi ,\psi ,\lambda)}} =
\normord{\int \mathcal{D} \phi \,
e^{-\frac{c}{2}\phi^2-
\frac{1}{\lambda}\bar{\psi}\left(\partial- c \phi\right)\psi}} \\
& S(\phi ,\bar\psi ,\psi ,\lambda) = -\frac{c}{2}
\int_{y}^{x} \phi^2(t) dt
-\frac{1}{\lambda}
\int_{y}^{x} \bar{\psi}(t)\left(\partial_t- c\phi(t)\right)\psi(t) dt.
\end{align*}
Using this formula and equation~\eqref{NN}, we rewrite the
product of the $\mathrm{Q}$-operators in the normal ordered form
\begin{align*}
\mathbb{Q}^{x}_{y}(\lambda)
\mathbb{Q}^{x}_{y}(\mu) = \int \mathcal{D} \phi_1
\int \mathcal{D} \phi_2 \,e^{-\frac{c}{2}\left(\phi_1^2+\phi_2^2\right)}
\normord{e^{\bar{\psi}\left(
\frac{1}{\lambda\mu}\left(\partial- c\phi_1\right)
\left(\partial- c\phi_2\right)-
\frac{1}{\lambda}\left(\partial- c\phi_1\right)-
\frac{1}{\mu}\left(\partial- c\phi_2\right)\right)\psi}}
\end{align*}
Clearly, the last expression is symmetric under the permutation $\lambda \rightleftarrows \mu$.

So, we have shown in two ways, by taking the continuum limit
as $\ell \to \infty$ of the known $\mathrm{Q}$-operator for
the XXX spin chain and directly in the continuous
model, that the $\mathrm{Q}$-operator for
the quantum NLS model is given by expression~\eqref{QZ}.
It will be interesting to study a connection between the
$\mathrm{Q}$-operator and the Backlund transformation~\cite{S3}
for this model, hopefully we will return to it in future.

\section*{\Large Appendix}
\addcontentsline{toc}{section}{Appendix}

\renewcommand{\theequation}{\Alph{section}.\arabic{equation}}
\renewcommand{\thetable}{\Alph{table}}
\setcounter{section}{0}
\setcounter{table}{0}

\appendix

\section{Normal symbols of operators}

All formulas in the Appendix are written in the
universal notation. The general formula for a
product of two normal ordered operators can be
expressed in two forms: using functional derivatives
\begin{align*}
\normord{F_1(\bar{\psi} ,\psi)} \,
\normord{F_2(\bar{\psi} ,\psi)} =
\left.\normord{e^{\frac{\delta^2}{\delta\phi\delta\bar{\phi}}}
F_1(\bar{\psi} ,\phi)F_2(\bar{\phi} ,\psi)}
\right|_{\substack{\phi=\psi \\ \bar{\phi}=\bar{\psi}}}
\end{align*}
or a functional integral
\begin{align}\label{FI}
\normord{F_1(\bar{\psi} ,\psi)} \,
\normord{F_2(\bar{\psi} ,\psi)} =
\normord{\int \mathcal{D} \phi \mathcal{D}
\bar{\phi} \, F_1(\bar{\psi} ,\phi) F_2(\bar{\phi} ,\psi)  e^{-(\phi-\psi)(\bar{\phi}-\bar{\psi})}}
\end{align}
Applying any of these formulas it is 
not difficult to show that
\begin{align}\label{a}
e^{a\bar{\psi}}  e^{\bar{a}\psi} \normord{F(\bar{\psi} ,\psi)} =
\normord{e^{a\bar{\psi}+\bar{a}\psi} F(\bar{\psi}+\bar{a} ,\psi)}
\ \ \ \,,\ \ \  
\normord{F(\bar{\psi} ,\psi)} \, e^{a\bar{\psi}}  e^{\bar{a}\psi}=
\normord{e^{a\bar{\psi}+\bar{a}\psi} F(\bar{\psi} ,\psi+a)}
\end{align}
Similarly, some general relations
\begin{align}
\label{NN}
\normord{e^{\bar{\psi}D_1\psi}} \,
\normord{e^{\bar{\psi}D_2\psi}}
& =
\normord{e^{\bar{\psi}\left(D_1D_2+D_1+D_2\right)\psi}}
\\
\label{b1}
\normord{e^{\alpha \bar{\psi}\partial\psi +
\beta \left(\bar{\psi}\psi\right)^2 + \gamma \bar{\psi}\psi}}
& = \normord{e^{\gamma \bar{\psi}\psi}} \,
\normord{e^{\frac{\alpha}{\gamma+1}\bar{\psi}\partial\psi + \frac{\beta}{(\gamma+1)^2}\left(\bar{\psi}\psi\right)^2}} \\
\nonumber
\left(1+\lambda\right)^{\bar{\psi}\psi}
& =
\normord{e^{\lambda\bar{\psi}\psi}}
\end{align}
can be proved using formula~\eqref{FI}.
In the first formula $D_1$, $D_2$ are any
linear operators acting on the fields $\psi(x)$ and $\bar{\psi}(x)$.

By putting in~\eqref{b1} $\alpha = -\frac{1}{\lambda}$, $\beta = \frac{ab}{2\lambda^2}$
and $\gamma = \pm\frac{ab}{\lambda}$, we obtain formulas for the
$\mathrm{Q}$-operator
\begin{align}\nonumber
\normord{e^{-\frac{1}{\lambda}\bar{\psi}\partial\psi +
\frac{ab}{2\lambda^2}\left(\bar{\psi}\psi\right)^2
\pm\frac{ab}{\lambda}\bar{\psi}\psi}} & =
\normord{e^{\pm\frac{ab}{\lambda}\,\bar{\psi}\psi}}\,
\normord{e^{-\frac{1}{\lambda\pm ab}\bar{\psi}\partial\psi + \frac{ab}{2(\lambda\pm ab)^2}\left(\bar{\psi}\psi\right)^2}}
\\
\label{b3}
& = \left(1\pm\frac{ab}{\lambda}\right)^{\bar{\psi}\psi}\,
\normord{e^{-\frac{1}{\lambda\pm ab}\bar{\psi}\partial\psi + \frac{ab}{2(\lambda\pm ab)^2}\left(\bar{\psi}\psi\right)^2}}
\end{align}

\section{Ordered exponentials}

A fundamental solution of the differential
equation
\begin{align}\label{def1}
\partial_x F(x,y) = U(x) F(x,y),
\quad \left.F(x ,y)\right|_{x=y} = \II
\end{align}
can be expressed with the help of the ordered
exponential
\begin{align}\label{Pexp}
F(x ,y) = \overset{\curvearrowleft}{\exp}
\int_{y}^{x} U(t) d t =
\II + \int_{y}^{x} d t\, U(t) +
\int_{y}^{x}d t_1\,\int_{y}^{t_1}d t_2\, U(t_1)  U(t_2) + \ldots
\end{align}
The main formulas with ordered exponentials
\begin{align}\label{1}
& F(x ,y) F(y ,z) = F(x ,z); \quad F(y ,x) = F^{-1}(x ,y)\\
\label{2}
& F_{V}(x ,y) = F(x ,y) + \int_y^x d z F(x ,z) V(z) F(z ,y) \\
\nonumber
& + \int_{y}^{x} d z_1\,\int_{y}^{z_1}d z_2\,
F(x ,z_1) V(z_1)  F(z_1 ,z_2) V(z_2) F(z_2 ,y)+\ldots
\end{align}
The last one defines perturbation series for a solution of the
equation~\eqref{def1} with the slightly changed matrix
$U(x)\to U(x)+V(x)$; here $F(x, y)$ is the solution of the
original equation~\eqref{def1} with the matrix $U(x)$.
In order to derive this result, consider an equation for
the function $F_{V}(x, y)$
\begin{align*}
\partial_x F_{V}(x ,y) =
\left(U(x)+V(x)\right) F_{V}(x ,y),
\quad \left.F_{V}(x ,y)\right|_{x=y} = \II.
\end{align*}
Let us look for a solution in the form $F_{V}(x,y) =
F(x,y)f(x,y)$. Then for $f(x,y)$ we get
\begin{align*}
\partial_x f(x ,y) =
F^{-1}(x ,y) V(x) F(x ,y) f(x ,y),
\quad \left.f(x\,,y)\right|_{x=y} = \II,
\end{align*}
hence the solution can be expressed in terms of~\eqref{Pexp}
substituting $U(x)\to F^{-1}(x,y) V(x) F(x,y)$. Using this and formula~\eqref{1}, 
we get
\begin{align*}
f(x ,y) =
\II + \int_{y}^{x} d t\, F(y ,t) V(t) F(t ,y) +
\int_{y}^{x}d t_1\,\int_{y}^{t_1}d t_2\,
F(y ,t_1) V(t_1) F(t_1 ,t_2) V(t_2) F(t_2 ,y)
+ \ldots
\end{align*}
Multiplying both sides by $F(x,y)$, we obtain formula~\eqref{2}.
Note also that same formula~\eqref{2} could be used to calculate
the variation $\delta F(x,y) = F_{\delta U}(x,y) - F(x,y)$. Substituting
$U\to U+\delta U$,
\begin{gather}\label{3}
\delta F(x ,y) = \int_y^x d z F(x ,z) \delta U(z) F(z ,y) \\
\nonumber
+ \int_{y}^{x} d z_1\,\int_{y}^{z_1}d z_2\,
F(x ,z_1) \delta U(z_1)  F(z_1 ,z_2) \delta U(z_2) F(z_2 ,y)+\ldots
\end{gather}
In this work we are interested in the variations of a form
$\delta U = \phi(x) A$, where $\phi(x)$ is a scalar function and
a matrix $A$ does not depend on $x$. The general formula~\eqref{3}
leads to the explicit expressions for the derivatives
\begin{align}\label{var1}
& \frac{\delta F(x ,y)}{\delta \phi(z)} =
F(x ,z) A F(z ,y), \quad y<z<x \\
\label{var2}
& \frac{\delta^2 F(x ,y)}{\delta \phi(z_1)\delta \phi(z_2)} =
F(x ,z_1) A F(z_1 ,z_2) A F(z_2 ,y), \quad y<z_2<z_1<x,
\end{align}
and so on.

\subsection{Gauge transformations}

Suppose we want to represent a solution
of the differential equation
\begin{align*}
\partial_x F(x ,y) = U(x) F(x ,y),
\quad \left.F(x ,y)\right|_{x=y} = \II
\end{align*}
in the form $F(x,y) = S(x) G(x, y) S^{-1}(y)$, where
$S(x)$ is some known matrix. One could easily
derive the equation for $G(x,y)$
\begin{align*}
\partial_x G(x ,y) = \left(S^{-1}(x) U(x) S(x)-S^{-1}(x) \partial_x S(x)\right) G(x ,y),
\quad\left.G(x ,y)\right|_{x=y} = \II.
\end{align*}
Using ordered exponentials, we can rewrite these formulas
as
\begin{align*}
\overset{\curvearrowleft}{\exp}
\int_{y}^{x}\left(S^{-1}(t) U(t) S(t)-S^{-1}(t) \partial_t S(t)\right)d t =
S^{-1}(x) \left(\overset{\curvearrowleft}{\exp}
\int_{y}^{x}U(t)d t\right) S(y).
\end{align*}

\section{Commutation relations for the monodromy
matrix in the continuous model}\label{C}

In this Appendix we prove relation~\eqref{glob}.
Let us remark that this formula was proved by
E. K. Sklyanin in his paper~\cite{S1} in two ways.
In fact, our proof repeats the main steps of E. K.
Sklyanin's proof, but with more emphasis on
functional methods, and, of course, is of purely
methodical interest.

The proof is based on explicit formulas for
products of normal ordered operators.
Let us rewrite the product of the monodromy matrices
on the left-hand side of the relation~\eqref{glob} in
the normal ordered form
\begin{align}\nonumber
& \mathbb{T}_1(\lambda) \mathbb{T}_2(\mu) = \normord{T_1(\lambda)}\,
\normord{T_2(\mu)} =
\left.\normord{e^{\frac{\delta^2}{\delta\phi\delta\bar{\phi}}}
T_1(\bar{\psi} ,\phi)T_2(\bar{\phi} ,\psi)}
\right|_{\substack{\phi=\psi \\ \bar{\phi}=\bar{\psi}}} \\
\nonumber
& = \sum_{n=0}^{\infty} \frac{1}{n!} \int \prod_{k=1}^n d z_k
\left.\normord{\frac{\delta^n T_1(\bar{\psi} ,\phi)}
{\delta\phi(z_1)\ldots\delta\phi(z_n)}
\frac{\delta^n  T_2(\bar{\phi} ,\psi)}{\delta\bar{\phi}(z_1)\ldots\delta\bar{\phi}(z_n)}}
\right|_{\substack{\phi=\psi \\ \bar{\phi}=\bar{\psi}}}\\
\label{PT}
& = \sum_{n=0}^{\infty} \int_{Z_n} \prod_{k=1}^n d z_k
\left.\normord{\frac{\delta^n T_1(\bar{\psi} ,\phi)}
{\delta\phi(z_1)\ldots\delta\phi(z_n)}
\frac{\delta^n  T_2(\bar{\phi} ,\psi)}{\delta\bar{\phi}(z_1)\ldots\delta\bar{\phi}(z_n)}}
\right|_{\substack{\phi=\psi \\ \bar{\phi}=\bar{\psi}}}
\end{align}
Note that expression~\eqref{PT} is symmetric under the permutation of
the variables $z_k$, so we get rid of the factor $\frac{1}{n!}$ by passing to
integration over the fundamental domain $Z_n: y < z_n < \ldots < z_1 < x$.

In order to compute functional derivatives, let us substitute
$\delta U(x) = \phi(x) a\sigma_{+} + \bar{\phi}(x) b\sigma_{-}$
in~\eqref{3}. Hence we obtain the formulas similar to~\eqref{var1},~\eqref{var2}
\begin{align*}
& \left.\frac{\delta T_1(\bar{\psi} ,\phi)}
{\delta\phi(z)}
\right|_{\phi=\psi} =
T^{x}_{z}(\lambda) a\sigma_{+}
 T^{z}_{y}(\lambda)\otimes\II , \\
& \left.\frac{\delta T_2(\bar{\phi} ,\psi)}
{\delta\bar{\phi}(z)}
\right|_{\bar{\phi}=\bar{\psi}} =
\II\otimes T^{x}_{z}(\mu) b\sigma_{-} T^{z}_{y}(\mu).
\end{align*}
Then we rewrite the product of these expressions in
the following form
\begin{align*}
\left.\frac{\delta T_1(\bar{\psi} ,\phi)}
{\delta\phi(z)}
\frac{\delta  T_2(\bar{\phi} ,\psi)}{\delta\bar{\phi}(z)}
\right|_{\substack{\phi=\psi \\ \bar{\phi}=\bar{\psi}}} =
T^{x}_{z}(\lambda ,\mu)
 \left(a\sigma_{+}\otimes b\sigma_{-}\right)
 T^{z}_{y}(\lambda ,\mu),
\end{align*}
where $T^{x}_{z}(\lambda, \mu) =
T^{x}_{z}(\lambda)\otimes T^{x}_{z}(\mu)$.
In the same way we compute the higher derivatives,
for example,
\begin{align*}
\left.\frac{\delta^2 T_1(\bar{\psi} ,\phi)}
{\delta\phi(z_1)\delta\phi(z_2)}
\frac{\delta^2  T_2(\bar{\phi} ,\psi)}{\delta\bar{\phi}(z_1)\delta\bar{\phi}(z_2)}
\right|_{\substack{\phi=\psi \\ \bar{\phi}=\bar{\psi}}} =
T^{x}_{z_1}(\lambda ,\mu)
 \left(a\sigma_{+}\otimes b\sigma_{-}\right) T^{z_1}_{z_2}(\lambda ,\mu)
 \left(a\sigma_{+}\otimes b\sigma_{-}\right)
T^{z_2}_{y}(\lambda ,\mu).
\end{align*}
In fact, the sum in expression~\eqref{PT} has the same form
as the perturbation series~\eqref{2} with substitutions
$F(x,y)\to T^{x}_{y}(\lambda,\mu)$ and $V(x) \to a\sigma_{+}\otimes b\sigma_{-}$.
The product $T^{x}_{y}(\lambda,\mu) = T^{x}_{y}(\lambda)\otimes T^{x}_{y}(\mu)$
is a solution of the equation
\begin{align*}
\partial_x T^{x}_{y}(\lambda ,\mu) =
\left(L(x ,\lambda)\otimes\II +\II\otimes L(x ,\mu)\right) T^{x}_{y}(\lambda ,\mu)
\end{align*}
with initial condition $T^{x}_{x}(\lambda,\mu) = \II$. However
applying the normal ordering operation, we obtain the
perturbation series~\eqref{2} for $T^{\prime x}_{y}(\lambda,\mu)$,
where $T^{\prime x}_{y}(\lambda,\mu)$ is a solution of the equation
\begin{align*}
\partial_x T^{\prime x}_{  y}(\lambda ,\mu) =
\left(L(x ,\lambda)\otimes\II +\II\otimes L(x ,\mu)+ab \sigma_{-}\otimes\sigma_{+}\right)
T^{\prime x}_{  y}(\lambda ,\mu),
\end{align*}
with initial condition $T^{\prime x}_{x}(\lambda,\mu) = \II$.
Thus
\begin{align*}
\mathbb{T}_1(\lambda) \mathbb{T}_2(\mu) =
\left.\normord{e^{\frac{\delta^2}{\delta\phi\delta\bar{\phi}}}
T_1(\bar{\psi} ,\phi)T_2(\bar{\phi} ,\psi)}
\right|_{\substack{\phi=\psi \\ \bar{\phi}=\bar{\psi}}} =
\normord{T^{\prime x}_{  y}(\lambda ,\mu)}
\end{align*}
Similarly, we derive the formula for the inverse product
of the monodromy matrices
\begin{align*}
\mathbb{T}_2(\mu)
\mathbb{T}_1(\lambda) =
\left.\normord{e^{\frac{\delta^2}{\delta\phi\delta\bar{\phi}}}
T_2(\bar{\psi} ,\phi)T_1(\bar{\phi} ,\psi)}
\right|_{\substack{\phi=\psi \\ \bar{\phi}=\bar{\psi}}} =
\normord{T^{\prime\prime x}_{    y}(\mu ,\lambda)}
\end{align*}
where $T^{\prime\prime x}_{y}(\mu,\lambda)$ solves the equation
\begin{align*}
\partial_x T^{\prime\prime x}_{    y}(\mu ,\lambda) =
\left(L(x ,\lambda)\otimes\II +\II\otimes L(x ,\mu)+ab \sigma_{+}\otimes\sigma_{-}\right)
T^{\prime\prime x}_{    y}(\mu ,\lambda)
\end{align*}
with initial condition $T^{\prime\prime x}_{x}(\mu,\lambda) = \II$.
So the global relation~\eqref{glob} corresponds to the local one
\begin{align*}
& \mathbb{R}_{12}(\lambda-\mu) \left(L(x ,\lambda)\otimes\II +\II\otimes L(x ,\mu)+ab \sigma_{-}\otimes\sigma_{+}\right) \\
\nonumber
& = \left(L(x ,\lambda)\otimes\II +\II\otimes L(x ,\mu)+ab \sigma_{+}\otimes\sigma_{-}\right)
\mathbb{R}_{12}(\lambda-\mu),
\end{align*}
which can be easily proved.
Note also that taking into account the explicit formula
for the $\mathrm{R}$-matrix~\eqref{R}, it actually
remains to check the simpler relation
\begin{align*}
& \mathbb{R}_{12}(\lambda-\mu)
\left(\nicefrac{\lambda}{2} \sigma_3\otimes\II +
\nicefrac{\mu}{2} \II\otimes\sigma_3 +ab \sigma_{-}\otimes\sigma_{+}\right) \\
\nonumber
& = \left(\nicefrac{\lambda}{2} \sigma_3\otimes\II +
\nicefrac{\mu}{2} \II\otimes\sigma_3+ab \sigma_{+}\otimes\sigma_{-}\right)
\mathbb{R}_{12}(\lambda-\mu).
\end{align*}


\end{document}